\documentclass{emulateapj}
\usepackage{apjfonts}

\shorttitle{Intracluster Short Gamma-Ray Bursts}
\shortauthors{Niino \& Totani}

\begin{document}

\title{Intracluster Short Gamma-Ray Bursts by Compact Binary Mergers}

\author{Yuu Niino\altaffilmark{1,2} and Tomonori Totani\altaffilmark{1}}
\altaffiltext{1}{Department of Astronomy, School of Science, Kyoto University,
Sakyo-ku, Kyoto 606-8502, Japan}
\altaffiltext{2}{niinou@kusastro.kyoto-u.ac.jp}

\begin{abstract}
One of the possible origins of short gamma-ray bursts (SGRBs) is merging
of compact binaries, and the effect of large kick velocity is a
signature that can be used as an observational test for this
hypothesis. Intracluster SGRBs that escaped from a host galaxy in a
galaxy cluster are interesting in this context, since they would escape
more easily by cluster tidal force, and would have brighter afterglow
luminosity by dense intracluster gas, than those in general field
galaxies. Here we calculate the escape fraction of compact binaries from
their host galaxies in a galaxy cluster, and discuss some observational
implications. We found that the escape fraction strongly depends on the
nature of dark matter subhalos associated with member galaxies.  If the
amount of dark matter around member galaxies is not large and the
gravitational potential for an escaping binary is determined mostly by
stellar mass, most of SGRBs should escape and be observed as hostless,
which is a much higher fraction than those in the field. Hence,
statistics of intracluster SGRBs could give important information about
the dark matter distribution in galaxy clusters, as well as hints for
the origin of SGRBs.
\end{abstract}


\section{INTRODUCTION}

Short gamma-ray bursts (SGRBs) occur not only in star forming galaxies
but also in early type galaxies, indicating a longer time scale of the
delay from star formation to the SGRB events \citep{geh05,vil05,ber05b}
than that of the long gamma-ray bursts (LGRBs) that are associated with
massive star collapses [see \citet{pir04};
\citet{nak07} and \citet {lee07} for reviews]. A strong candidate for
the origin of SGRBs satisfying this nature is merging of compact object
binaries (e.g., double neutron star binaries or neutron star -
black hole binaries).

Compact binaries are expected to have large systemic velocities by the
kick given at the time of compact object formation, as inferred from the
large proper motion of pulsars. Because of the long time scale and the
large velocity, SGRBs are expected to occur with a large offset from the
centers of galaxies compared with stellar distribution.  There are some
model calculations of this effect (Bulik, Belczy\'nski, \& Zbijewski
1999; Bloom, Sigurdsson, \& Pols 1999; Fryer, Woosley,\& Hartmann 1999)
and such an expectation is indeed consistent with observations, though
the statistics is still limited \citep{lee05,fox05,blo07}. However,  
truly intergalactic SGRBs that are ejected from
their host galaxies would be difficult to identify, because of the
difficulty to identify their host galaxies. Furthermore, we do not
expect a strong afterglow emission in low density environment in the
intergalactic field, making the identification even more difficult [see
also \citet{blo06}; \citet{nak07}].

SGRBs in galaxy clusters are particularly interesting in this context,
since compact binaries would be more efficiently stripped away from
their host galaxies and become intracluster SGRBs, because of the strong
tidal force in the cluster gravity field. The idea of tidal stripping 
is consistent with observations of intracluster stars 
\citep{vil94,oka02,dur02,gal03,ger07}. We expect that such
intracluster SGRBs could have brighter afterglows because of the denser
environment of the intracluster gas than that in the intergalactic
field, which would help the accurate determination of SGRB locations. 
A few SGRBs have already been observed in galaxy clusters, though the
sample is still small \citep{ber07,shi07}. The statistics of such
intracluster SGRBs in the future data would then give us some
information about the origin of SGRBs, 
and the gravitational potential of galaxies in a cluster. 

Here we calculate the expected event rate of such intracluster SGRBs,
assuming that SGRBs are indeed the products of compact binary
mergers. We also discuss possible implications for future observations.
We assume $\Lambda$CDM cosmology with $\Omega_\Lambda=0.7$,
$\Omega_M=0.3$, and $H_0=70\ \mathrm{km\ s^{-1}\ Mpc^{-1}}$.

\section{FORMULATIONS}
\label{sec:formulation}

\subsection{Models of a Galaxy Cluster and Its Member Galaxies
at $z=0$}
\label{sec:model}

First we construct a model of a galaxy cluster at $z=0$, and
will consider its evolution in the next subsection.
As a typical galaxy cluster, we consider a spherically symmetric cluster
with a total cluster mass of $10^{15} M_\odot$, a virial radius of
$R_{\rm vir}= 1\ \mathrm{Mpc}$, and the Navarro-Frenk-White (NFW,
Navarro, Frenk, \& White 1995) density profile for dark matter (DM) with
a concentration parameter of $c_{\rm halo}=5$.  All these parameters are
typical for a rich cluster [e.g., \citet{sch06}]. We assume the
Schechter function for the luminosity function of member galaxies, with
the shape parameters (i.e., $L_*$ and $\alpha$) independent of radius
from the cluster center, $R$.  We use the estimate of the Schechter
parameters by the Sloan Digital Sky Survey data for galaxy
clusters \citep[Table 2]{han05}. The $R$-dependence of the normalization
of the luminosity function, $\phi_*$, is assumed to be the NFW profile,
but with the concentration parameter $c_{\rm galaxy}=1.2$ that is
different from that for the dark matter \citep{han05}.

We then determine the stellar mass and density profile within a member
galaxy as follows. It is well known that early-type galaxies are the
dominant population in rich galaxy clusters \citep {oem74,tho06},
and we assume all cluster galaxies to be elliptical galaxies for the
simplicity. The observed luminosity function is in the $r'$ band, and we
translate it into the $B$ band by a typical color of elliptical
galaxies, $B-r' =-1.4$. We assume the ratio of the stellar mass to light
to be $9.8M_\odot/L_\odot$ in the $B$ band. These quantities are from
the model used by \citet{tot00}, which can reproduce the observed galaxy
properties well. Stellar density profile within member galaxies is
assumed to be the Hernquist profile \citep{her90}, with the
characteristic radius parameter chosen so that a half of the total mass
is included within the observed half-light radius, $r_{\rm half}$, 
which is calculated by the power-law relation to $B$ luminosity fitted
to observations \citep{tot00}.

On the other hand, the density profile of dark matter associated with
the cluster member galaxies as substructure in the whole cluster is
uncertain and poorly known. Especially, the dark matter subhalos
extending to the outer region of member galaxies are expected to be
vulnerable to the tidal forces by galaxy interactions and/or the overall
cluster gravitational potential. Here we consider the two extreme
cases: (i) there is no significant DM substructure or subhalos
associated with member galaxies, and the gravitational potential well of
galaxies is determined simply by the stellar mass profile (the no
subhalo case), and (ii) the DM subhalos are associated with member
galaxies with a similar amount to the field galaxies (the preserved
subhalo case). The amount and nature of subhalos depend on the cluster forming
processes, and probably the reality is between these two extreme cases.

For the preserved subhalo case, we calculate the virial mass ($M_ {\rm
vir}$) of the subhalos from the stellar mass of a galaxy and the
universal ratio of the dark-to-baryonic matter $\Omega_M/\Omega_b=5.9$
\citep{spe03}. The virial radius ($r_{\rm vir}$) of the subhalo is
calculated from $M_{\rm vir}$ and the 1-dimensional central velocity
dispersion ($\sigma_v$) of the galaxy, as $GM_{\rm vir}/2r_{\rm vir}=
3\sigma_v^2$.  Here, the velocity dispersion is that for stars
calculated from the galaxy luminosity using the Faber-Jackson relation
\citep{vau82}. An almost similar velocity dispersion is obtained also
from the stellar mass and $r_{\rm half}$ calculated above and assuming
the virial relation, giving a consistency check for our treatment.  We
assume the NFW profile for the DM subhalos, and the concentration
parameter is calculated by the formula given by \citet{bul01} for
subhalos included in larger virialized halos, which are based on
cosmological N-body simulations: $c \simeq c_* (M_{\rm vir}/M_*)^
\Gamma$ at $z=0$ where $c_* = 7$, $\Gamma = -0.3$, and
$M_*\simeq1.5\times10^ {13} h^{-1}M_\odot$.

\subsection{Cluster and Galaxy Evolution}
\label{section:evolution}
It is well known that most stars in elliptical galaxies formed at
high redshift ($z > 1$) and they evolved passively to the present
time (e.g., Yamada et al. 2005). Major mergers could change drastically
the stellar mass distribution and gravitational potential of member
galaxies, but a recent numerical simulations by \citet{mur07} indicates
that the majority of member galaxies in a cluster do not undergo major
mergers, except for the brightest central galaxy which is formed in the
collision of many galaxies. Therefore we make a reasonable
assumption that the stellar mass distribution does not evolve in
cluster member galaxies.

Though galaxies form at the early epoch of $z > 1$, the establishment
of the overall cluster potential should be significantly later according
to the standard picture of hierarchical structure formation.  We
estimate the epoch of cluster formation using the extended
Press-Schechter approximation \citep{lac93}. It predicts that about half
of mass of a $10^{15} M_\odot$ cluster at $z=0$ is already included in
the largest progenitor at $z_F \sim 0.45$. Therefore we assume that the
tidal force by the cluster potential starts to affect member galaxies at
this redshift. We do not take into account the evolution of the cluster
potential at $z < z_F$, and this is a reasonable approximation  
because the
time scale of cluster potential evolution is much larger than the
orbital period of compact binaries in member galaxies.  The evolution of
DM subhalos is difficult to predict without detailed numerical
simulations, and we simply apply the above two extreme models with no
evolution, which would cover the realistic evolution.

\subsection{Escape of Compact Object Binaries}
\label{section:escape} In a galaxy with a given luminosity, we can
calculate the orbit of compact object binaries in the gravitational
potential as modeled in \S \ref{sec:model}, if the initial location
and velocity are given.  We calculate the initial velocity by the sum of
the original stellar velocity $\mathbf{v}_{\rm org}$ at the location and
the kick velocity $\mathbf{v}_{\rm kick}$ given when the compact objects
are formed. The Maxwell distribution having the 1-dimensional velocity
dispersion $\sigma_v$ defined in the previous section is assumed for
$\mathbf{v}_{\rm org}$. The direction of both $\mathbf{v}_{\rm org}$ and
$\mathbf{v}_{\rm kick}$ are assumed to be isotropic and random.

Although the velocity distribution of observed single pulsars is well
fitted by a Gaussian \citep{hob05}, that for compact object binaries is
not well known.  A few observed binary pulsars have bulk motion
velocities of $\sim$ 100--200 km/s \citep
{wex00,ran04,wil04}.  \citet{bul99}, \citet{blo99} and
\citet{fry99} theoretically estimated the velocities of compact binaries
that remain gravitationally bound after supernova explosions, to be
several hundreds km/s.  We assume the distribution of $\mathbf{v}_{\rm
kick}$ to be a single isotropic Gaussian, i.e., each 1-dimensional
component of $\mathbf{v}_{\rm kick}$ is a Gaussian with the standard
deviation $\sigma_k$.  We calculate the cases of two different values of
$\sigma_k$ = 100 and 300 km/s.  The mean (the standard deviation) of the
corresponding $|\mathbf{v}_{\rm kick}|$ distribution then becomes 160
(66) and 480 (200) km/s for $\sigma_k$ = 100 and 300 $\mathrm{km/s}$,
respectively.

We solve the motion of binaries until their merger time (the time from a
compact binary formation to its merger by gravitational wave
radiation).  Though the merger time generally ranges more than three
orders of magnitude, $10^{7-10}$ yr (e.g., Tutukov \& Yungelson 1994;
Bulik et al. 1999), we are interested in SGRBs in galaxy clusters. Most
of galaxies in clusters have formed their stars at high redshift ($z
\gtrsim 2$), and observed SGRBs are typically at $z \sim
0.2$. Therefore, the time between these epochs, i.e., $\sim 10^{10}$ is
appropriate for the merger time in this work.

We consider that a binary has escaped from its host galaxy once its
distance from the host galaxy center becomes larger than the tidal
radius $r_{\rm tide}$ of the host galaxy after the formation of the
galaxy cluster, i.e., $z < z_F$. Here, the tidal radius is defined as
the radius where the tidal force by the overall cluster potential is the
same as the binding force in the host galaxy. The tidal radius is
numerically calculated for a given set of galaxy luminosity and location
in the cluster. We then calculate the mean escape fraction as a
function of $R$, taking a weighted average over the host galaxy
luminosity, initial location in the host, and kick velocity.

It should be noted that some fraction of stars are distributed at $r >
r_{\rm tide}$ with the assumed stellar mass profile in host
galaxies. Such stars would be stripped from host galaxies and become
intracluster stars. Compact object binaries in such stellar populations
would all contribute to the intracluster SGRBs.  We find that this
fraction is about 2 \% in the preserved subhalo case and 10 \% in the no
subhalo case. It seems that the no subhalo case is preferred (see
\ref{section:obs}), from a comparison of these values with the various
observational estimates of the abundance of intracluster stars in galaxy
clusters.

\section{RESULTS}

\begin{figure}
\includegraphics{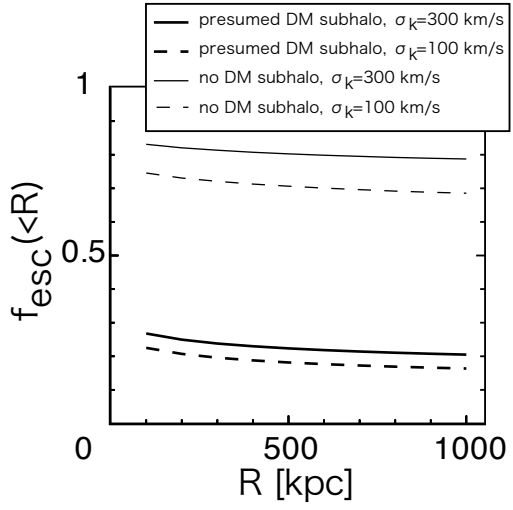}
\caption
{The escape fraction of compact binaries from host galaxies in a
galaxy cluster averaged within $R$, where $R$ is  the distance
from the center of the galaxy cluster. (The virial radius of the cluster
is 1 Mpc.)
The models with no dark matter subhalos is shown by thin lines,
while the models with preserved subhalos are shown by thick lines.
The solid and dashed lines are for
different values of the standard deviation of the kick
velocity distribution, $\sigma_k=300$ and 100 km/s, respectively.}
\label{fig:graph}
\end{figure}

The results are shown in Figure \ref{fig:graph}, where the mean escape
fraction within a given radius from the cluster center, $f_{\rm esc}(<
R)$, is plotted. We find that the escape fraction largely depends on
the existence of the DM substructure; the escape fraction is modest with
$f_{\rm esc} \sim 0.2$ in the preserved subhalo case, while most
binaries will be ejected in the no subhalo case.  The dependence on the
radius from the cluster center or on the kick velocity is not as
significant as the effect of subhalos within the parameter ranges
investigated.

For comparison, we calculate the case of field galaxies not in
clusters. Again, we only consider elliptical galaxies, and their
properties are calculated in the same way for a given luminosity.
\citet{bul01} found that isolated halos have a different relation
between the virial mass and the concentration parameter from that for
subhalos, and we adopt $c_* = 9$ and $ \Gamma = -0.13$ here based on
their results.  The escape fraction averaged over galaxy luminosity is
simply calculated without taking into account the tidal force by
external gravity field. Here we use the luminosity function shape
parameters of \citet{bla01} for field galaxies; though these are derived
for all types of field galaxies, the luminosity function for each galaxy
type is rather uncertain.  The results are shown in Table 1, and we
found that the escape fraction in the field is not much different from
that in galaxy clusters in the preserved subhalo case, while a large
enhancement of the cluster $f_{\rm esc}$ is predicted in the case of no
subhalos.

\section{DISCUSSION}
\subsection{Detectability of SGRB Afterglows in Clusters}

Detection of afterglows is necessary to locate a GRB accurately enough
with respect to a host galaxy.  We discuss here the detectability of a
typical SGRB afterglow in the intracluster medium following the standard
afterglow model of \citet{sar98}. We simply use this isotropic model
without jet structure by using the isotropic equivalent total energy.
The jet break may reduce the expected flux at a later time compared with
the calculation here, but this crude estimate is sufficient here for our
purpose.

The isotropic-equivalent total energy in gamma-rays of SGRBs is
distributed in a wide range of $E_{\rm \gamma, iso} \sim
10^{49}$--$10^{51}$ erg, and the total initial kinetic energy of the
external shock ($E_{\rm iso}$) is expected to be similar \citep
{fox05,sod06}.  According to \citet{pan01}, we adopt the following
parameters: the fraction of energy density in magnetic field $
\epsilon_B=10^ {-2.4}$ and that in nonthermal electrons
$\epsilon_e=10^{-1.2}$; although these are for LGRBs, the values
inferred for available SGRBs are not much different.  We also assume the
power index of the luminosity decay, $\alpha = -1$ where $F_\nu \propto
t^{\alpha}$.  The typical particle density of intracluster medium is $n
\sim 10^ {-3} \ \rm cm^{-3}$ within a few hundreds kpc from the cluster
center (e.g., Lewis, Buote, \& Stocke 2003). We assume a typical
distance for SGRBs, $z = 0.2$.

Then the model predicts the expected flux at the observed frequency of
$\nu$ as $F_\nu \sim 1.5\ E_{50}^{4/3} \ n_{-3}^{1/2} \ \nu_{15}^{-2/3}
(t/ 10^3 \ \rm s)^{-1}\ \mu$Jy, where $E_{50} \equiv E_{\rm iso} /
(10^{50} \ \rm erg)$, $n_{-3} \equiv n/(10^{-3} \ \rm cm^{-3})$ and
$\nu_{15} \equiv \nu / (10^{15} \ \rm Hz)$.  In the X-ray band (1 keV),
the typical flux is then $\nu F_\nu \sim 9.3 \times 10^{-14} \ \rm erg \
cm^{-2} s^{-1}$ at $t=10^3$ s, which can be detected by existing X-ray
satellites (e.g., Gehrels et al. 2005).  {\it Swift} XRT can locate
afterglows with an accuracy of a few arcsec, and this is reasonably
accurate to discuss the association of an afterglow with a galaxy at
$z\sim0.2$.  In the optical ($R$) band, this flux corresponds to
$\sim$26 mag (AB) at $t=10^4$ s. Most afterglows with this level of
brightness have been missed in the past and current GRB follow up
observations. However, afterglows of brightest SGRBs ($E_{\rm iso} \sim
10^{51}$) may be detectable.

On the other hand, SGRBs ejected far from their host galaxies in the
normal field would occur in much lower-density environment.  The typical
density in general intergalactic medium would be estimated as $n \sim
\rho_c \Omega_b / m_p \sim10^{-7} \ \mathrm{cm}^ {-3}$, where $\rho_c$
and $m_p$ are the critical density of the universe and the proton mass,
respectively. The expected afterglow flux of an intergalactic SGRB is then more than one
order of magnitude fainter than those of the faintest SGRB afterglows
ever observed, such as GRB 050509B \citep{geh05} and GRB 050911
\citep[upper limit only]{pag06}. Therefore it seems difficult to detect
an afterglow of such an event.

\subsection{Comparison with Observations}
\label{section:obs}

\citet{ber07} examined all 16 SGRBs that were followed up by X-ray
observations with XRT of {\it Swift} or {\it Chandra}, and found that
three SGRBs are likely to be
associated with galaxy clusters, suggesting that the fraction of SGRBs
in galaxy clusters is about 20 \%.  Considering the statistical
uncertainty, it is consistent with the fraction of all stellar mass in
the universe bound in galaxy clusters ($\sim10$ \%, Fukugita, Hogen,
\& Peebles 1998).

Among the three SGRBs in galaxy clusters discussed in \citet{ber07}, GRB
050509B is apparently associated with the likely host galaxy at the
cluster redshift of $z = 0.226$.  The offset of GRB 050509B from its
host is 21--56 kpc, corresponding to 6--16$r_{\rm half}$.  The galaxy
cluster that contains the host is composed of two subclusters, and the
host galaxy is located at the center of the minor subcluster that is
about 270 kpc away from the center of the major subcluster \citep
{geh05,blo06}.  The location of GRB 050911 is in a cluster of $z =
0.165$, but its afterglow was too faint to associate it with any
particular galaxy \citep{pag06}.  GRB 050813 has three candidate host
galaxies near its {\it Swift} XRT location, and the galaxies belong to
two different galaxy clusters at $z=0.72$ and $z=1.8$
\citep{ber05a,ber06}.  Clearly, the current sample is too small to
derive any implications from a comparison with our results.
Future satellites for GRB study might detect SGRBs more efficiently
leading to a much larger sample of SGRBs.

Observations of intracluster diffuse light, stars,
type Ia supernovae, and planetary nebulae
indicate that some stars in a galaxy cluster are in intracluster medium,
perhaps removed from member galaxies \citep
{vil94,oka02,dur02,gal03,ger07}.  The fraction of intracluster
stars in all stars in a cluster is uncertain, but observational
estimates are typically $\sim 5$--20 \%.  These fractions are consistent
with our estimate of stars outside the tidal radius (see \S
\ref{section:escape}).  On the other hand, $f_{\rm esc}$ of SGRBs could
be much higher than these, up to $\sim$ 80 \% depending on the model
parameters.  If such a higher fraction of intracluster SGRBs than that
of intracluster stars is observed in the future, it would indicate the
effect of kick velocities on compact binaries, giving a further support
for the compact binary hypothesis of SGRBs. Note that this test is
difficult in the intergalactic field, since we do not know the fraction
of intergalactic stars and afterglows of intergalactic SGRBs are
difficult to detect.

\section{CONCLUSIONS}

We investigated the escape of compact binaries from their host galaxies
in galaxy clusters, which is enhanced by the tidal force of the cluster
gravity compared with general fields. We found that the escape
probability heavily depends on the uncertain distribution of subhalos associated with
member galaxies.  If the DM substructure has been destroyed by
interactions in a galaxy cluster and the escape of a binary is
determined mainly by gravity of stellar mass, most of compact binaries
in galaxy clusters should escape and become hostless intracluster
SGRBs. On the other hand, if the DM subhalos are associated to member
galaxies with a similar amount to field galaxies, the enhancement of
escape fraction is only modest compared with field galaxies: about 20 \%
for clusters while $\sim$ 10 \% for field galaxies.

Though the current observed data set is not sufficient to be
compared quantitatively with our results,
statistics of SGRB association with cluster galaxies in the future
data will give us important information for the
dark matter distribution in clusters, intracluster stars
in clusters, as well as the origin of SGRBs.

\acknowledgments
We would like to thank an anonymous referee for useful comments.
This work was supported by the Grant-in-Aid for the 21st Century COE
"Center for Diversity and Universality in Physics"
from the Ministry of Education, Culture,
Sports, Science and Technology (MEXT) of Japan.

\clearpage

\begin{table}
\begin{center}
\caption{escape fractions in galaxy cluster $f_{\rm esc}(<R_{\rm
vir})$ and in field \label{tb:F}}
\begin{tabular}{crrr}
\tableline\tableline
$\sigma_k$ (km/s) & Cluster (i) & Cluster (ii) & Field \\
\tableline
300 &0.79 &0.20 &0.10\\
100 &0.69 &0.16 &0.09\\
\tableline
\end{tabular}
\tablecomments
{The columns labeled as Cluster (i) and (ii) give the escape fractions
corresponding to (i) the no DM subhalo case
and (ii) the preserved DM subhalo case discussed
in \S \ref{sec:model}.}
\end{center}
\end{table}

\clearpage

\end{document}